\documentclass[sigconf]{acmart}

\AtBeginDocument{%
  \providecommand\BibTeX{{%
    \normalfont B\kern-0.5em{\scshape i\kern-0.25em b}\kern-0.8em\TeX}}}

\setcopyright{acmcopyright}
\copyrightyear{2019}
\acmYear{2019}

\acmConference[Anchorage '19]{2nd KDD Workshop on Anomaly Detection in Finance}{August 5, 2019}{Anchorage, Alaska}
\acmPrice{15.00}
\acmISBN{978-1-4503-9999-9/18/06}

\usepackage{bm} 
\usepackage{multirow}
\usepackage{booktabs} 
\usepackage{graphicx}
\usepackage{stfloats}
\graphicspath{{figures/}}
\usepackage{subfig}

\begin{document}

\title{Risk Management via Anomaly Circumvent: Mnemonic Deep Learning for Midterm Stock Prediction}

\author{Xinyi Li}
 \authornote{Both authors contributed equally to this research.}
\email{xl2717@columbia.edu}
\affiliation{%
  \institution{Columbia University}
}

\author{Yinchuan Li}
\email{liyinchuan@bit.edu.cn}
\orcid{0000-0002-4263-5130}
\authornotemark[1]
\affiliation{%
  \institution{Beijing Institute of Technology}
}

\author{Xiao-Yang Liu}
\email{xl2427@columbia.edu}
\affiliation{%
  \institution{Columbia University}
}

\author{Christina Dan Wang}
\email{christina.wang@nyu.edu}
\affiliation{%
  \institution{New York University Shanghai}
}








\renewcommand{\shortauthors}{Li and Li, et al.}

\begin{abstract}
  Midterm stock price prediction is crucial for value investments in the stock market. However, most deep learning models are essentially short-term and applying them to midterm predictions encounters large cumulative errors because they cannot avoid anomalies.
  In this paper, we propose a novel deep neural network Mid-LSTM for midterm stock prediction, which incorporates the market trend as hidden states. First, based on the autoregressive moving average model (ARMA), a midterm ARMA is formulated by taking into consideration both hidden states and the capital asset pricing model. Then, a midterm LSTM-based deep neural network is designed, which consists of three components: LSTM, hidden Markov model and linear regression networks. The proposed Mid-LSTM can avoid anomalies to reduce large prediction errors, and has good explanatory effects on the factors affecting stock prices. Extensive experiments on S\&P 500 stocks show that (i) the proposed Mid-LSTM achieves 2-4\% improvement in prediction accuracy, and (ii) in portfolio allocation investment, we achieve up to 120.16\% annual return and 2.99 average Sharpe ratio.
\end{abstract}




\maketitle

\section{Introduction}

 Midterm stock prediction is fundamental and important for investment companies and quantitative analysts. Relative to the high-risk short-term investment and the slow return long-term investment, the midterm investment has always been a matter of high concern. However, it has been proven to be a very difficult task since the stock market is a highly nonlinear dynamic system. In particular, the movements of stocks are influenced by immense factors such as interest rates, inflation rates, trader's expectation, catastrophe, political events and economic environments \cite{chang2009integrating}. \textcolor{black}{Hence, the stock price is noisy and contain many anomalies. If the anomalies cannot be effectively detected, the risk of midterm predictions will be very high.}
 

The recurrent neural networks (RNNs) is widely used to predict stock prices~\cite{kamijo1990stock,ding2015deep, zhuoran2018practical,li2019optimistic}. However, the main problem is that traditional RNNs cannot solve the long-term sequence dependency problem~\cite{sutskever2014sequence}. \textcolor{black}{Hence, these RNNs are unable to remember early anomalies and apply them in the midterm stock predictions.}
In~\cite{hochreiter1997long}, an effective method called long short-term memory (LSTM) is proposed to address the long time lag problem. Some stock sequence prediction methods using LSTM have been proposed~\cite{nelson2017stock,selvin2017stock,zhuge2017lstm}, which shows the applicability and potential of LSTM in stock prediction. In~\cite{nelson2017stock,selvin2017stock}, the method of using LSTM to predict short-term returns and stock prices is analyzed. In~\cite{zhuge2017lstm}, the LSTM neural network is combined with emotional analysis to predict the short-term stock prices. Unfortunately, the traditional LSTM network has limited ability to predict short-term stocks because short-term stock prices are noisy and unstable. The distribution of financial time series varies over time, which means that stock prices are non-stationary and inherently complex. In contrast, LSTM is more suitable for midterm prediction since it has the memory cell and can retain the pattern of the sequence. The main difficulty of the midterm stock prices prediction is that it requires the full sequence prediction, i.e., the latter predicted prices are based on the previous predicted prices. \textcolor{black}{Hence, the prediction errors will accumulate, if the previous predicted prices are biased, the final prediction is very inaccurate, which may lead to higher investment risks.}


\textcolor{black}{
In order to reduce investment risks, a suitable neural network should be used to obtain an accurate predicted price. Moreover, an appropriate stock price model should consider the effects of anomalies. For value investors, they need a risk management model to effectively detect trends in mid-term stock price changes, thereby effectively avoiding investment uncertainty caused by anomalies.
The capital asset pricing model (CAPM)~\cite{merton1973intertemporal}, which describes the relationship between systematic risk and expected return of assets, is widely used for mid-term investments. Mid-term investments can effectively avoid the effects of anomalies by considering market factors. In addition, in recent years, the hidden Markov model (HMM)~ \cite{seymore1999learning} has been proven to have a good ability to detect hidden states. The HMM model can be used to detect hidden states in markets and trading volumes, etc., thereby effectively avoiding anomalies.
}

\textcolor{black}{This paper aims at predicting the mid-term stock prices with the minimum risk neural network.}
Considering that a single stock model cannot fully incorporate the impacts of various factors on prices, a stock price model that incorporates multiple models is proposed. First, the autoregressive moving average (ARMA)~\cite{box2015time} model, which is one of the most widely used linear models in time series prediction, is used to set up the stock midterm prediction problem. Then, a few variables that are highly related to stock price are considered in the midterm ARMA-based prediction model (Mid-ARMA), for example, the volume, whose changes may be a precursor to price changes. \textcolor{black}{Moreover, inspired by the CAPM, the market price and the correlation coefficient between the stock and market are considered in the Mid-ARMA model to improve the prediction accuracy.} 
Based on the Mid-ARMA model, a midterm LSTM-based deep neural network (Mid-LSTM) is proposed to predict stock prices, where the LSTM network is used to first predict the stock price, volume and market price. \textcolor{black}{Then, the HMM, which is a powerful statistical machine learning technique in information extraction tasks, is used to extract the hidden states between the stock price and volume.} Once the hidden states and correlation coefficient are available, the linear regression, which is an approach to model the relationship between a dependent variable and one or more independent variables, is used to cooperate the above variables to refine the predicted stock price. Finally, we compare the Mid-LSTM network with many traditional machine learning methods based on the S\&P 500 data. Various prediction results show that the Mid-LSTM network can obtain accurate predicted prices under different market conditions, which significantly improves the investment return. \textcolor{black}{Especially when the market is in an abnormal state, our proposed method is obviously superior to the traditional methods.}

The remainder of the paper is organized as follows. In Section 2, we introduce traditional stock price models and present the Mid-ARMA model for midterm stock prediction. In Section 3, we develop the Mid-LSTM deep neural network and present the training details. Data preprocessing, portfolio allocation methods and investment returns are provided in Section 4. Section 5 concludes the paper.

\section{Stock Price Model}
In this section, we first introduce traditional stock price models, and then present the proposed Mid-ARMA model for midterm stock prediction and the corresponding loss functions.

\subsection{Traditional Stock Price Models}

The ARMA model is one of the most widely used linear models for stock price prediction, where the future value is assumed as a linear combination of the past values and past errors~\cite{box2015time}. Let ${X}_t^\text{A}$ be the variable based on ARMA at time $t$, then we have
\begin{align}
\label{ARIMA}
    {X}_t^\text{A}=&~ f_1(\{X_{t-i}\}_{i=1}^p),\nonumber \\
    =&~ \mu + \sum_{i=1}^{p}\phi_i X_{t-i} - \sum_{i=1}^{q}\psi_j\epsilon_{t-j}+\epsilon_t,
\end{align}
where $X_{t-i}$ denotes the past value at time $t-i$; $\epsilon_{t}$ denotes the random error at time $t$; $\phi_i$ and $\psi_j$ are the coefficients; $\mu$ is a constant; $p$ and $q$ are integers that are often referred to as autoregressive and moving average polynomials, respectively.

Besides the past stock prices, many variables also have impact on stock price movements, such as the volume and market index. Volume refers to the number of shares or contracts traded during a given period. It has been used to infer whether an event had informational content and whether investors' interpretations of the information were similar or different~\cite{beaver1968information}. The empirical analyses in \cite{ying1966stock} and \cite{crouch1970volume} have three findings:

\begin{itemize}
    \item A small volume is usually accompanied by a fall in price;
    \item A large volume is usually accompanied by a rise in price;
    \item A large increase in volume is usually accompanied by either a large rise in price or a large fall in price.
\end{itemize}
The above observations imply that there are some hidden states $S_t$ between the stock price $X_t$ and volume $V_t$, i.e., 
\begin{align}
S_{t-i} = g(X_{t-i},V_{t-i}),~~i=1,...,p,
\end{align}
where $g(\cdot)$ denotes the hidden function. \textcolor{black}{Hidden variables are usually defined as variables that profoundly affect stock prices, but are invisible and need to be extracted from visible variables.}

The market index can be regarded as a hidden driving force and can be considered in the stock price model to reduce \textcolor{black}{the impact of anomalies}.
According to the CAPM~\cite{merton1973intertemporal}, which describes the relationship of risks and returns between stocks and the market, the expected return in equilibrium of asset $R$ is given by
\begin{align}
\label{CAPM1}
    \mathbb{E}(R)=r_f +\beta_{M}[\mathbb{E}(R_M)-r_f],
\end{align}
where $\mathbb{E}(R_M)$ denotes the expected market return; $r_f$ denotes the risk-free rate; and
\begin{align}
\label{CAPM2}
   \beta_{M}=\frac{\text{cov}(R,R_M)}{\textcolor{black}{\sigma}^2(R_M)}
\end{align}
is the Market Beta, with $\text{cov}(\cdot)$ being the covariance and $\text{$\sigma$}(\cdot)$ being the \textcolor{black}{standard variance}. 

\subsection{Mid-ARMA Model for Stock Price}

Stock price movements are affected by many factors, a single stock model cannot fully consider the impact of various factors on prices, resulting in higher prediction errors. Here we introduce multiple factors to jointly describe stock prices \textcolor{black}{and reduce the impact of anomalies}. First, the volume is added to the model. Then, the CAPM inspired us to incorporate the impact of market into the midterm stock prediction. To this end, our proposed Mid-ARMA model is given by
\begin{align}
\label{ARIMA_SPY_V}
	\hat{X}_t =&~ \alpha {X}_t^{\text{A}} + \underbrace{ c + \rho(\lambda {M}_t^{\text{A}} + \eta) }_{\text{CAPM-based}} + \underbrace{ \gamma g({X}_t^{\text{A}}, {V}_t^{\text{A}}) }_{\text{hidden states}},\nonumber \\
	=&~ \alpha {X}_t^{\text{A}} + \lambda \rho {M}_t^{\text{A}} + \eta \rho + \gamma {S}_t^{\text{A}} + c,
\end{align}
where $\alpha$, $\eta$, $\lambda$ and $\gamma$ are weighting factors; $c$ is a constant; the correlation coefficient{\footnote{Here we use the correlation coefficient instead of the Market Beta because we want to increase the impact of stock prices to expand the difference between different correlation coefficients.}} is calculated by
\begin{align}
\label{rho}
\rho = \frac{\text{cov}({X}_t^{\text{A}},{M}_t^{\text{A}})}{\textcolor{black}{\sigma}({X}_t^{\text{A}})\cdot\textcolor{black}{\sigma}({M}_t^{\text{A}})},
\end{align}
and the market index{\footnote{Since our model focuses on stock price prediction instead of stock return prediction, we use the market index here instead of the expected market return $\mathbb{E}(R_M)$.}} $M_t^{\text{A}}$ and volume $V_t^{\text{A}}$ are also predicted based on the ARMA model similarly as in \eqref{ARIMA} as
\begin{align}
\label{ARMA-M}
    {M}_t^{\text{A}}=&~ f_2(\{M_{t-i}\}_{i=1}^p), \\
\label{ARMA-V}
    {V}_t^{\text{A}}=&~ f_3(\{V_{t-i}\}_{i=1}^p).
\end{align}

\begin{figure}[!tb]
	\centering
	
	\subfloat{\includegraphics[width=2.8in]{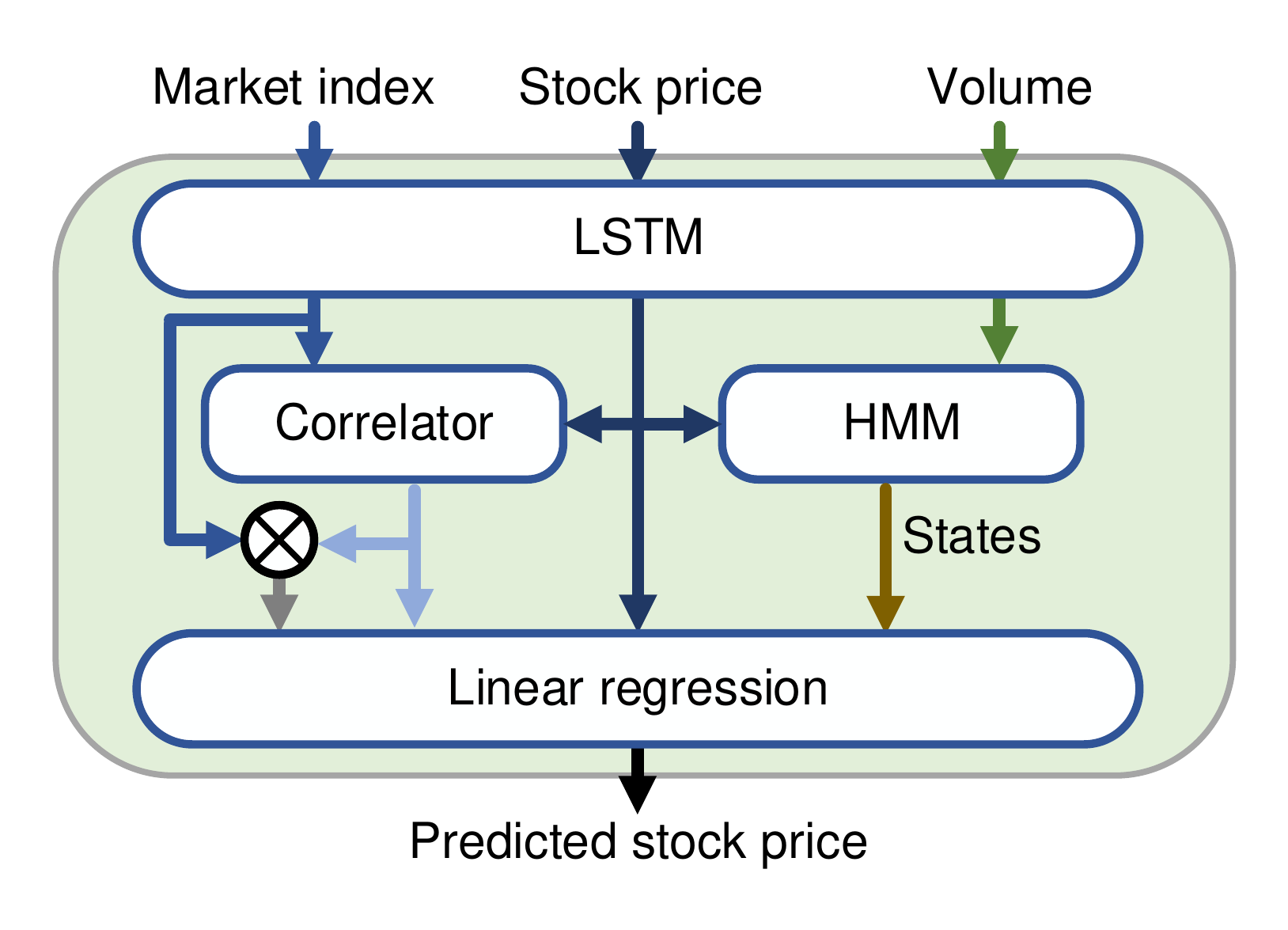}}
	
	\caption{Mid-LSTM framework.}
	\label{figure:Mid-LSTM}
\end{figure}

\subsection{Loss Function for Training Mid-LSTM}

Our proposed Mid-LSTM neural network aims at predicting the midterm stock price based on the Mid-ARMA model by minimizing the following loss function:
\begin{align}
\label{L2}
    {\mathcal{L}}_{1} = \min\sum_{t=p+1}^{p+T} \left \| X_t- \hat{X}_t \right \|_2^2,
\end{align}
\textcolor{black}{where $T$ denotes the number of prediction time slots, i.e., $t = 1,...,p$ are the observations (training input data), $t = p+1,...,p+T$ are the predicts (training output data)}; $\hat X_t$ is given in \eqref{ARIMA_SPY_V}. Obtaining $\hat X_t$ obviously  requires first predicting the stock price, market index as well as volume, by minimizing the loss function:
\begin{align}
\label{L1}
    {\mathcal{L}}_{2} =&~ \min \sum_{t=p+1}^{p+T} \left\| {\mathcal{Y}}_t^{\text{train}} - f({\mathcal{X}}_t^{\text{train}})  \right\|_2^2,
\end{align}
where ${\mathcal X}_t^{\text{train}}$ and ${\mathcal Y}_t^{\text{train}}$ are respectively the training input set and training output set:
\begin{align}
{\mathcal X}_t^{\text{train}} =&~ \{(X_{t-i},M_{t-i},V_{t-i})\}_{i=1}^{p}, \\
{\mathcal Y}_t^{\text{train}} =&~ (X_t,M_t,V_t).
\end{align}
\textcolor{black}{The Mid-LSTM loss functions are designed according to the Mid-ARMA. The LSTM network (with loss function $L_1$) is according to the ARMA model, while the linear regression (with loss function $L_2$) in Mid-LSTM is designed according to Mid-ARMA model.}

\section{Design of Mid-LSTM Neural Network}

In this section, our Mid-LSTM deep neural network is first presented, which consists of three components: LSTM, hidden Markov model and linear regression network. The ``min-max'' normalization method and the Mid-LSTM training details are then provided.

\subsection{Overview of Our Mid-LSTM Scheme}

\textcolor{black}{The Mid-LSTM aims at predicting the midterm stock price by minimizing the loss functions in \eqref{L2} and \eqref{L1}. The Mid-LSTM focus on the midterm prediction of stocks (30-60 days).} The scheme of our Mid-LSTM deep neural network is shown in Figure~\ref{figure:Mid-LSTM}. First, the LSTM neural network is used to obtain the predicted stock price, market index and volume. Afterwords, the predicted hidden states $S_t$ between the predicted volume and stock price are determined by the HMM, and the predicted correlation coefficient $\rho$ is calculated \textcolor{black}{by the correlator based on \eqref{rho}.} Finally, the linear regression model is used to refine the predicted stock price, ${\hat X}_t$ is obtained based on the Mid-ARMA model by minimizing the loss function in \eqref{L2}. The Mid-LSTM incorporates the hidden states between the volume and stock price, and incorporates the CAPM model to control the midterm investment risk. Hence, Mid-LSTM has good performance for predicting the midterm stock price, especially for those stocks that are highly correlated with the market.


\begin{figure}
	\centering
	
	\subfloat{\includegraphics[width=3.4in]{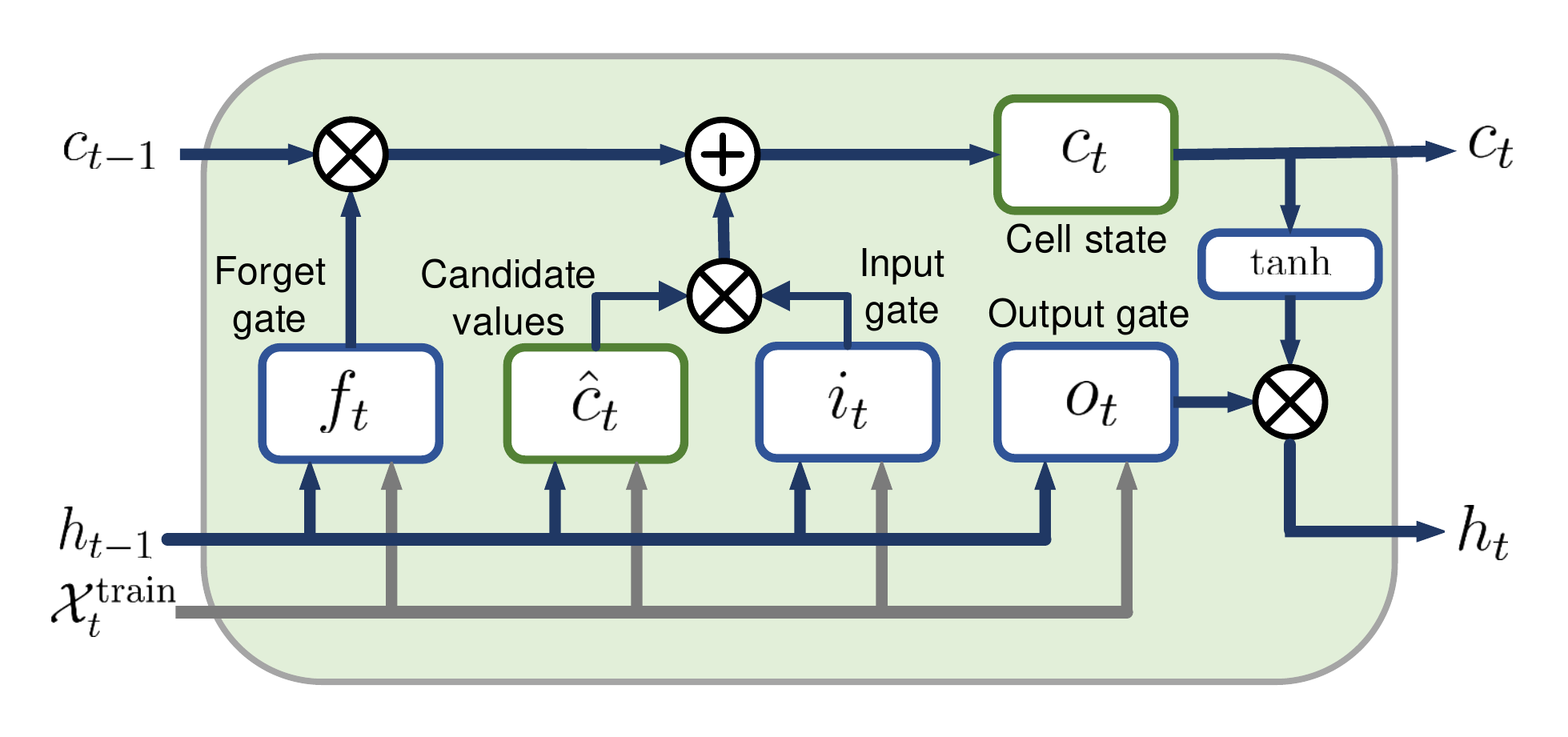}}
	
	\caption{Framework of LSTM in Figure~\ref{figure:Mid-LSTM}.}
	\label{figure:LSTM}
\end{figure}

\subsubsection{LSTM Network}

\textcolor{black}{The Mid-LSTM is designed according to the Mid-ARMA.} The LSTM network in Figure~\ref{figure:Mid-LSTM} consists of an input layer, one or more hidden layers and an output layer. The main feature of LSTM networks is that the hidden layer consists of memory cells. Each memory cell has a core recurrently self-connected linear unit called ``Constant Error Carousel (CEC)''\cite{gers2002learning}, which provides short-term memory storage and has three gates (see Figure~\ref{figure:LSTM}):
\begin{itemize}

\item Input gate, which controls the information from a new input to the memory cell, is given by
\begin{align}
\label{LSTM1_1}
   i_t =&~ \sigma(W_i\times[h_{t-1},{\mathcal X}_t^{\text{train}}]+b_i), \\
\label{LSTM1_2}
   \hat{c}_t = &~\tanh(W_c\times[h_{t-1},{\mathcal X}_t^{\text{train}}]+b_c),
\end{align}
where $h_{t-1}$ is the hidden state at the time step $t-1$; $i_t$ is the output of the input gate layer at the time step $t$; $\hat{c}_t$ is the candidate value to be added to the output at the time step $t$; $b_i$ and $b_c$ are biases of the input gate layer and the candidate value computation, respectively; $W_i$ and $W_c$ are weights of the input gate and the candidate value computation, respectively; and sigmoid $\sigma(x) = 1/(1+e^{-x})$ is a pointwise nonlinear activation function.

\item Forget gate, which controls the limit up to which a value is saved in the memory, is given by
\begin{align}
\label{LSTM2_1}
   f_t = \sigma(W_f\times[h_{t-1},{\mathcal X}_t^{\text{train}}]+b_f),
\end{align}
where $f_t$ is the forget state at the time step $t$, $W_f$ is the weight of forget gate; and $b_f$ is the bias of forget gate.

\item Output gate, which controls the information output from the memory cell, is given by
\begin{align}
\label{LSTM3_1}
   c_t =&~ f_t \times c_{t-1}+i_t \times \hat{c}_t,\\
   o_t =&~ \sigma(W_o\times[h_{t-1},{\mathcal X}_t^{\text{train}}]+b_o),\\
   h_t =&~ o_t \times \tanh(c_t),
\label{LSTM3_2}
\end{align}
where new cell states $c_t$ are calculated based on the results of the previous two steps; $o_t$ is the output at the time step $t$; $W_o$ is the weight of the output gate; and $b_o$ is the bias of the output gate\cite{qin2017dual}.

\end{itemize}

\begin{figure}
	\centering
	
	\subfloat{\includegraphics[width=2.5in]{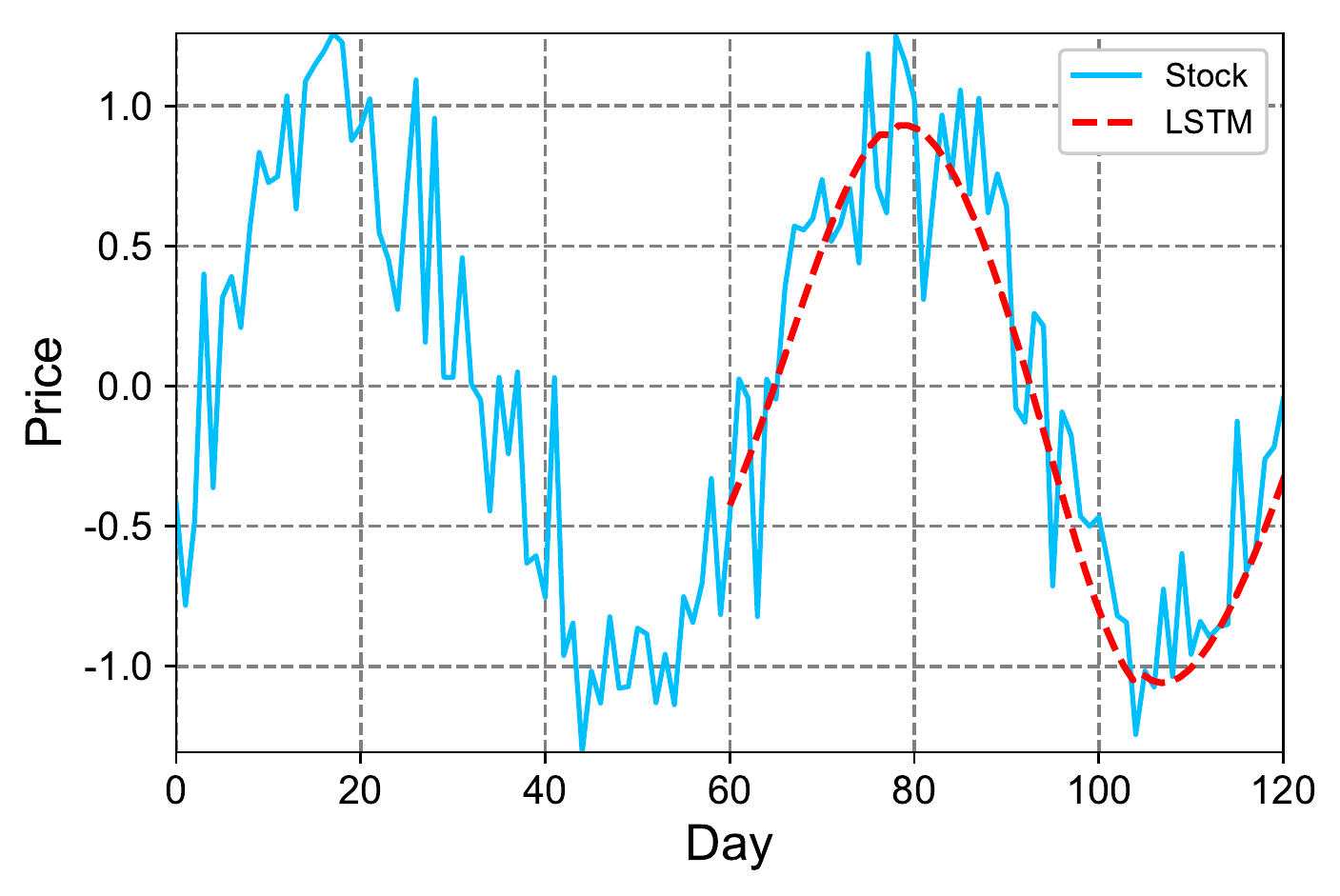}}
	
	\caption{\textcolor{black}{LSTM prediction results of a simulated stock. The stock price is generated by a sinusoidal function plus Gaussian white noise, where the sine function represents the midterm trend of the stock, and the Gaussian white noise represents the abnormal fluctuations of the stock. We can see that the LSTM prediction results circumvent the anomalies and are very consistent with the stock trend (sinusoidal function).}}
	\label{figure:LSTM-prediction}
\end{figure}

\textcolor{black}{The input gate, forget gate and output gate prevent memory content from being disturbed by anomalies (unrelated input and output), thus LSTM is suitable for learning sequences containing longer term patterns~\cite{malhotra2015long}. In contrast, short-term stock price trends are affected by many factors and are highly random, hence LSTM is not suitable for dealing with short-term stock trend prediction. A simply example is shown in Figure~\ref{figure:LSTM-prediction} to further illustrate this point, where LSTM predicts the simulated stock prices (60 to 120 days) based on the input prices (0 to 59 days). The stock trend is represented by a sinusoidal function, and then superimposed with Gaussian white noise to represent abnormal fluctuations in the stock price. We can see that LSTM can predict the mid-term stock trend quite well, while the short-term trends, which are highly correlated with the anomalies, are unpredictable (during 61-63 days, the stock price fell, but LSTM still predicts an upward trend).}

\subsubsection{HMM Network}
The HMM is a double stochastic process whose potential stochastic process is unobservable, but can be observed by another set of stochastic symbols~\cite{rabiner1986introduction}. A HMM assumes there is a Markov Chain, which might be entirely unobservable. However, there is an observation model, which, for each possible state of the Markov Chain, gives us a pre-distribution for the data that is observable. Hence, we use HMM to extract hidden states between the volume and price.

The HMM in Figure~\ref{figure:Mid-LSTM} consists of three parts as follow:
\begin{itemize}
    \item There are $K$ finite states in the model, within a state the signal processes some measurable and distinctive properties.
    \item At each time $t$, a new state is entered based on the transition probability distribution, which depends on the previous state.
    \item After each transition is made, an output symbol is produced according to the probability distribution, which depends on the current state. 
\end{itemize}

\subsubsection{Linear Regression Network}

\textcolor{black}{The Mid-ARMA defines the independent variables that are used in the linear regression.} The linear regression in Figure~\ref{figure:Mid-LSTM} is an approach to model the relationship between a dependent variable and one or more independent variables, which can determine the unknown weighting factors $\alpha$, $\eta$, $\lambda$, $\gamma$ and $c$ in \eqref{ARIMA_SPY_V}. Linear regression is an efficient supervised learning algorithm for prediction problems. It finds the target variable by finding the most appropriate fit line between the independent variables and the dependent variable. \textcolor{black}{The main advantage is that the fitted line has the minimum error from all points.} It has good explanatory meanings, the assigned weights can help analysts to find the most influential hidden variables in the market.


\subsection{Mid-LSTM Network Training}

\subsubsection{Normalization Process}

To detect midterm stock price pattern, it is necessary to normalize the stock price data. Different stocks have different domains and scales. Data normalization is defined as adjusting values measured on different scales to a uniform scale~\cite{ hafezi2015bat}. \textcolor{black}{If we train the model without normalization, the model would not converge.}

Since Mid-LSTM requires the stock patterns during training, we use ``min-max'' normalization method to reform dataset, which keeps the pattern of the data, as follow:
\begin{align}
\label{N3}
 X_{t}^{n}=\frac{X_{t}-\min({X_{t}})}{\max({X_{t}})-\min({X_{t}})},
\end{align}
where $X_{t}^{n}$ denotes the data after normalization.

Accordingly, de-normalization is required at the end of the prediction process to get the original price, which is given by
\begin{align}
\label{N4}
 \hat{X}_{t}=\hat{X}_{t}^{n}[\max({X}_{t})-\min({X}_{t})]+\min({X}_{t}),
\end{align}
where $\hat{X}_{t}^{n}$ denotes the predicted data and $\hat{X}_{t}$ denotes the predicted data after de-normalization. \textcolor{black}{Similarly, the market index is also normalized, but the volume is not normalized, since the HMM does not require the normalization processing.}

\subsubsection{Network Training}

The LSTM network in Figure~\ref{figure:Mid-LSTM} has six layers \textcolor{black}{(an LSTM layer, a dropout layer, an LSTM layer, an LSTM layer, a dropout layer, a dense layer, respectively)}. The dropout layers (with dropout rate 0.2) prevent the network from overfitting. The dense layer is used to reshape the output. Since a network will be difficult to train if it contains a lot of LSTM layers~\cite{sak2014long}, we use three LSTM layers here. In each LSTM layer, the loss function is the mean square error (MSE). The Adam~\cite{kingma2014adam} is used as optimizer, since it is straightforward to implement, computationally efficient and well suited for problems with large data set and parameters.


The HMM is trained based on the stock price and volume, so the observation dimension is $2$. \textcolor{black}{The normalized stock price and real volume are used as inputs, which are continuous numbers but are discrete based on the number of days, i.e., for each day, the dimension is 2, and the training input lasts for a window size of $N=60$ days. For the training output (hidden state), there is only one state per day, so the output dimension is 1, and there are a total of 4 different hidden states for 60 days.} The number of training iterations is set as 10. The number of hidden states is set as 4, i.e., we simply consider there are $K = 4$ hidden states between the stock price and volume as follows:
\begin{itemize}
    \item Large trading volume, high stock price;
    \item Large trading volume, low stock price;
    \item Small trading volume, high stock price;
    \item Small trading volume, low stock price.
\end{itemize}

After the LSTM and HMM are well trained, we use the outputs of them as the inputs of the linear regression model, and training it by minimizing the loss function in \eqref{L2}.

\section{Performance Evaluation}

In this section, we validate our Mid-LSTM based on the S\&P 500 stocks. The rolling window based data preprocessing and portfolio allocation method are provided.

\subsection{Rolling Window Based Data Preprocessing}

\begin{figure}[!b]
	\centering
	\subfloat{\includegraphics[width=3.0in]{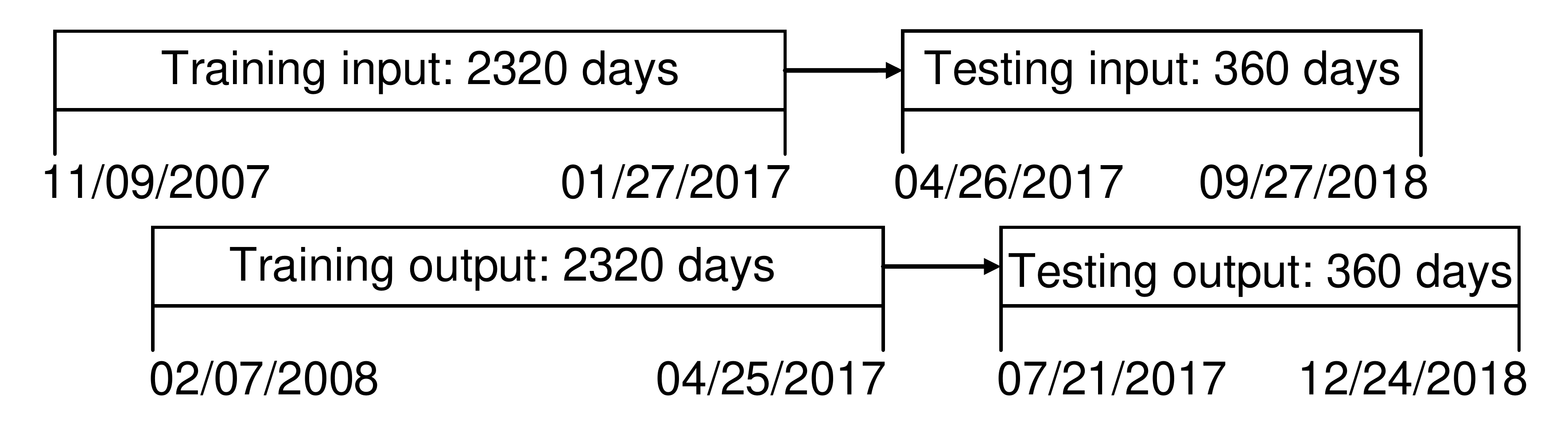}}
	\caption{Schematic diagram of rolling window.}
	\label{figure:rolling}
\end{figure}

\begin{figure*}[!htb]
	\centering
	\subfloat{\includegraphics[width=7.4in]{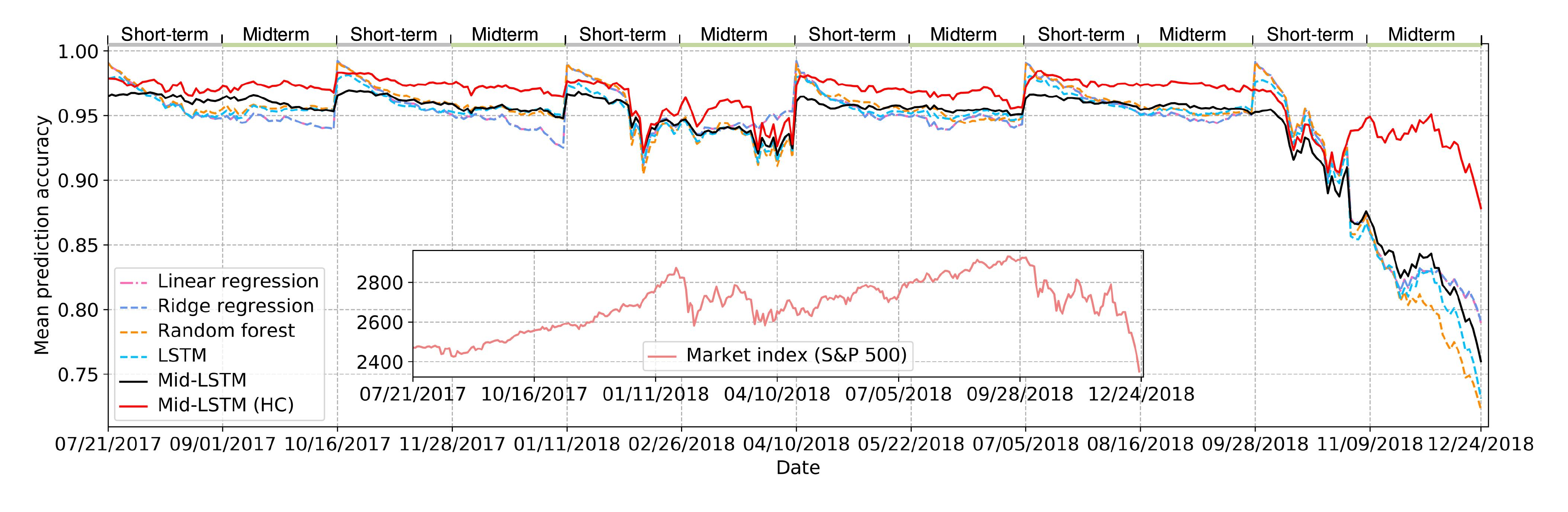}}
	\caption{Mean prediction accuracies of the Mid-LSTM and traditional methods. The prices between 07/21/2017 - 10/16/2017 are predicted based on real prices between 04/26/2017 - 07/20/2017, prices between 10/17/2017 - 01/11/2018 are predicted based on real prices between 07/21/2017 - 10/16/2017, and so forth. The predicted prices between 09/01/2017 - 10/16/2017, 11/28/2017 - 01/11/2018 and so on are defined as midterm prediction prices.}
	\label{figure:MPA-results}
\end{figure*}

The data for this project are the historical S\&P 500 component stocks, which are downloaded from the Yahoo Finance. The S\&P 500 is used as the market index. We use the data over the period of about 10 years (from 01/02/2009 to 12/24/2018). \textcolor{black}{ For other index to study (e.g., DJI (Dow 30), Nasdaq and Russell 2000), they are composed of a subset of stocks in our study or of similar stocks, so the average performance will be quite similar to the presented results.} The stocks with missing data are deleted, and the dataset we used eventually contains 451 stocks. Each stock records the close price and trading volume of 2,800 trading days.

About 85\% of the dataset (from 11/09/2007 to 04/25/2017 including 2380 trading days) is used as the training data, and the remaining dataset (from 04/26/2017 to 12/24/2018 including 420 trading days) is used as the testing data. A rolling window is used to separate data. We chose the window size of $N = 60$ days, which allows the neural network to get glimpses of the shape of the input sequence, and hence will hopefully teach itself to build up a pattern of the stock based on the prior window received. Then, according to the length of the window, the training data is divided into 2320 sets of training input data (each set length 59) and training output data (each set length 1). The testing data is divided into input and output data of 6 windows (see Figure~\ref{figure:rolling}).

We focus on the midterm prediction of stocks (30-60 days) in this project, which requires the full sequence prediction. Unlike the point-by-point prediction processing, which predicts the stock price all based on the input real stock price, the full sequence prediction processing predicts the midterm stock price based on the data that predicted in the previous predictions. In particular, the training window is first initialized with all real training data. Then we shift the window and add the first predicted point to the last point of training window to predict the next point and so forth. Once the input window consists entirely of past predicted points, we stop the prediction and let the window shift the entire window length forward, i.e., reset the window with the real training or testing data, and then start a new full sequence prediction again.


\subsection{Price Prediction Results}

We calculate the mean prediction accuracy (MPA) to evaluate the proposed methods, which is defined as 
\begin{align}
\label{accuracy1}
\text{MPA}_{t} = 1 - \frac{1}{L}\sum_{\ell=1}^{L} \frac{|X_{t,\ell}-\hat{X}_{t,\ell}|}{X_{t,\ell}},
\end{align}
where $X_{t,\ell}$ is the real stock price of the $\ell$-th stock on the $t$-th day, $L$ is the number of stocks and $\hat{X}_{t,\ell}$ is the corresponding prediction result. 

Figure~\ref{figure:MPA-results} shows the $\text{MPAs}$ of the proposed Mid-LSTM and some traditional methods (linear regression, ridge regression, random forest\footnote{\textcolor{black}{Note that we choose the random forests instead of the boosting methods (e.g., the  gradient  boosting  and  AdaBoost  algorithms) for comparison, since the boosting methods are sensitive to outliers and anomalies and not suitable for our prediction problem.}} and traditional LSTM). \textcolor{black}{Furthermore, based on the training data, we select 50 stocks that are most highly correlated (HC) with the market and then do prediction for comparison}. The Mid-LSTM (HC) in Figure~\ref{figure:MPA-results} represents the $\text{MPA}_{t}$ of 50 stocks that are highly correlated with the market. In Table~\ref{tab:MPA}, we give the mean MPA results for the midterm prices (09/01/2017 - 10/16/2017, 11/28/2017 - 01/11/2018 and so on), i.e., \textcolor{black}{since we are concerned with midterm predict accuracy, we only calculate the accuracy from day 30 to day 60 in each window.} From Figure~\ref{figure:MPA-results} and Table~\ref{tab:MPA} we can see that the predicted midterm prices of Mid-LSTM are more accurate than that of other traditional methods, especially for those high market-related stocks. This indicates that the linear regression model in Mid-LSTM well recognizes the hidden variables that affect stock prices and assigns them the appropriate weights.

\textcolor{black}{Note that the results are obtained by running many trials, since we train stocks separately and predict each price individually due to the different patterns of stock prices. This in total adds up to 451 runs. The results shown in Table~\ref{tab:MPA} is the average of these 451 runs. Furthermore, we provide results for 6 duration over a 10-year period in Figure~\ref{figure:MPA-results}. Over the long history, the performance of our algorithm is always better than the traditional ones. Especially when the market changes sharply (anomalies occur), the MPA of our Mid-LSTM is significantly higher than the traditional methods (see the duration 09/28/2018 - 12/24/2018 in Figure~\ref{figure:MPA-results}), which shows that our Mid-LSTM is robust. Based on the proposed Mid-ARMA model, the proposed Mid-LSTM neural network effectively avoids anomalies. The investment risk based on this prediction results is reduced.}




\color{black}Besides the MPA, we also evaluate the prediction trend accuracy (TA), which is defined as follows:
\begin{align}
\label{accuracy2}
\text{TA}=\frac{1}{6}(\sum_{w=1}^{6}\frac{1}{L}(\sum_{\ell=1}^{L}\text{flag}_{\ell,w})),
\end{align}
where
\begin{align}
\label{flag}
\text{flag}_{\ell,w}=\left\{
\begin{aligned}
1,~~~~  & \hat{X}_{0,w}\leq\hat{X}_{59,w}~\text{and}~X_{0,w} \leq X_{59,w},\\
1,~~~~  & \hat{X}_{0,w}\geq\hat{X}_{59,w}~\text{and}~X_{0,w} \geq X_{59,w},\\
0,~~~~  & others,
\end{aligned}
\right.
\end{align}
with $X_{0,w}$ and $X_{59,w}$ denoting the true stock prices of the first day and the last day in the $w$-th window, respectively; and $\hat{X}_{0,w}$ and $\hat{X}_{59,w}$ denoting the corresponding estimates.

\begin{table}
\centering
\begin{tabular}{ccc}  
\toprule
Method  & Mean MPA \\
\midrule
Linear regression    & 0.9253  \\
Ridge regression     & 0.9253  \\
Random forest        & 0.9235  \\
LSTM      & 0.9258  \\
\textbf{Mid-LSTM}  & \textbf{0.9308}  \\
\textbf{Mid-LSTM (HC)}  & \textbf{0.9637}\\
\bottomrule
\end{tabular}
\caption{Predicted Mean MPA results.}
\label{tab:MPA}
\end{table}

\begin{table}
\centering
\begin{tabular}{ccc}
\toprule
Method  & TA \\
\hline
Linear regression       & 0.8049      \\
Ridge regression        & 0.8071     \\
Random Forest           & 0.7650    \\
LSTM                    & 0.8160    \\
\textbf{Mid-LSTM}  & \textbf{0.8460}  \\
\textbf{Mid-LSTM (HC)}  & \textbf{0.9200}\\
\bottomrule
\end{tabular}
\caption{Predicted TA results}
\label{tab:trend}
\end{table}


\begin{table*}[!htb]\small
\centering
\begin{tabular}{c|ccccccc|ccccccc}
\toprule
\multirow{2}{*}{Method} &
\multicolumn{7}{c|}{Mean variance portfolio allocation (\%)} &
\multicolumn{7}{c}{Minimum variance portfolio allocation (\%)} \\
\cline{2-15}
  & R-1 & R-2 & R-3 & R-4 & R-5 & R-6 & Ave & R-1 & R-2 & R-3 & R-4 & R-5 & R-6 & Ave \\
\hline
Linear & 78.07  & 122.19  & 57.31  & 70.43  &80.92 &73.42  &80.39 &73.22 &115.94 &60.90 &62.73 &77.04 &65.39&75.87\\
Ridge  & 78.32  & 122.29  & 56.88  & 69.21  &80.73 &75.56  &80.50 &73.37 &115.81 &60.54 &59.07 &76.87 &67.64 &75.50\\
RF     & 81.72  & 102.07  & 59.14  & 68.69  &76.58 &119.07 &84.55 &78.15 &92.98 &53.09 &63.61 &69.85 &118.12 &79.30 \\
LSTM              & 69.11  & 96.62   & 65.09  & 79.31  &74.22 &-96.23 &48.02 &64.93 &87.31 &47.09 &77.85 &73.82 &-99.50 &41.92\\
\textbf{Mid-LSTM} & \textbf{92.18}  & \textbf{115.69} & \textbf{43.45} & \textbf{82.22} & \textbf{82.27} & \textbf{120.16} & \textbf{89.33} &\textbf{91.77} &\textbf{109.02} &\textbf{34.86} &\textbf{77.26} &\textbf{79.69} &\textbf{118.55} &\textbf{85.19}\\
\bottomrule
\end{tabular}
\caption{Portfolio returns of the first set of asset.}
\label{tab:cumu}
\end{table*}


\begin{table*}[!htb]\small
\centering
\begin{tabular}{c|ccccccc|ccccccc}
\toprule
\multirow{2}{*}{Method} &
\multicolumn{7}{c|}{Annualized portfolio return (\%)} &
\multicolumn{7}{c}{Annualized Sharpe ratio    \ \ (Risk-free rate: 1.5\%)} \\
\cline{2-15}
  & R-1 & R-2 & R-3 & R-4 & R-5 & R-6 & Ave & S-1 & S-2 & S-3 & S-4 & S-5 & S-6 & Ave \\
\hline
Linear &25.10 &20.91 &25.45  &25.23 &33.84 &14.63   &24.20 &2.69 &2.80 &1.33 &1.96 &3.56 &0.89 &2.21\\
Ridge  &25.13 &21.09 &24.75  &25.65 &33.19 &14.25   &24.01 &2.69 &2.82 &1.29 &2.00 &3.42 &0.87 &2.18\\
RF     &25.78 &32.53  &10.50 &22.58 &33.70 &14.01 &23.18 &3.80 &4.97 &0.46 &1.77 &4.40 &0.80 &2.70 \\
LSTM   &24.95 &27.98  &4.88 &31.68 &29.93 &26.47 &24.31 &3.51 &4.07 &0.18 &2.29 &3.80 &1.52 &2.56\\

\textbf{Mid-LSTM} & \textbf{37.50}  & \textbf{42.95} & \textbf{9.61} & \textbf{11.79} & \textbf{33.15} & \textbf{24.07} & \textbf{26.51} &\textbf{5.46} &\textbf{5.62} &\textbf{0.40} &\textbf{0.74} &\textbf{4.42} &\textbf{1.28} &\textbf{2.99}\\
\bottomrule
\end{tabular}
\caption{Portfolio returns and Sharpe ratios of the second set of asset based on mean variance portfolio allocation method.}
\label{tab:cumu2}
\end{table*}

We calculate the TA through all 6 testing windows. For traditional methods and Mid-LSTM, we calculate the trend accuracy of all stocks $(L=451)$. Moreover, we calculate the 50 stocks $(L=50)$ that are highly correlated with the market. From Table~\ref{tab:trend} we can see that the TA of the proposed Mid-LSTM is more accurate than that of traditional methods. The prediction TA of Mid-LSTM (HC) increases by approximately 12\% over traditional methods. We conclude that our Mid-LSTM produces more accurate results than traditional methods and the risk is low. Note that among the metrics, the TA can better measure the risk of the prediction methods because it will not be affected by stock price differences. In contrast, due to different stock prices, the MPA may be greatly affected by individual stocks and it is not objective enough.

Furthermore, it is noteworthy that for time series, linear models generally perform better when the signals are more random. Because any nonlinear model needs some trends or patterns in the signal to be utilized. Since the short-term stock is very noise and has no pattern can be utilized, the linear model nearly has the best performance in average (see Figure~\ref{figure:MPA-results}). In contrast, our proposed Mid-LSTM aims to solve the midterm stock prediction problem based on the proposed Mid-ARMA model, which is a nonlinear model (containing the product term in equation (5)). Hence, it is reasonable that our proposed method is not as good as others in the short-term.




\color{black}


\subsection{Portfolio Allocation and Return Results}

Portfolio allocation is important for investment strategies because it balances returns and risks by assigning weights to each asset~\cite{xing2018intelligent}.
We consider two sets of assets to evaluate the investment return based on our Mid-LSTM. To start with, the cumulative return is defined as
\begin{align}
\label{CR}
\hat{C}=\prod_{t=1}^{T-1}(1+\hat{R}_{t+1}),
\end{align}
where 
\begin{align}
\label{LR}
\hat{R}_{t+1} = \log({\hat{X}_{t+1}}/{\hat{X}_{t}}),~~t=1,...,T-1
\end{align}
is the log return.
In the first asset set we select stocks with cumulative return $\hat{C}>1.15$ from all 451 stocks. In the second asset set we select stocks with cumulative return $\hat{C}>1.05$ from 50 most market-related stocks. 
\textcolor{black}{Note that our proposed Mid-LSTM is only intended to predict midterm stock prices, since the short-term stock is noisy and the ability of LSTM to remember patterns cannot be used in short-term stock prediction. We can see that traditional methods (e.g., random forest) are more suitable for short-term predictions (see Figure~\ref{figure:MPA-results}). Therefore, when investing with Mid-LSTM, only the stock price predicted in the second month (30-60 days) is used, and the stock price in the first month (0-30 days) is replaced by the predicted price by random forest.}

 We provide two portfolio allocation methods for investors with different risk appetites. The first it to use the mean-variance optimization to allocate the stocks we choose, which is suitable for investors who prefer to get a higher Sharpe ratio (the return earned per unit volatility). In particular, the weights $\bm w = [w_1,...,w_L]^T$ of $L$ stocks are determined by
\begin{align}
\label{con_mv1}
\bm{\hat{w}} =&~ \arg\max_{\bm w} \frac{r_{p}(\bm w)-r_{f}} {\sigma_{p}(\bm w)},\\
\text{s.t.}&~~\sum_{\ell=1}^{L} w_{\ell} =1,~w_{\ell} \in [0,1],~{\ell}=1,...,L,\nonumber
\end{align}
recall that $r_f$ is the risk-free rate, which is set as 1.5\%. The annualized return of portfolio $r_p(\bm w)$ and the annualized variance of portfolio $\sigma_p(\bm w)$ are respectively calculated by
\begin{align}
\label{Mean}
r_p(\bm w) =&~ 252 \sum_{{\ell}=1}^{L}w_{\ell} \left( \frac{1}{T-1}\sum_{t=1}^{T-1}\hat{R}_{t+1} \right), \\
\label{Var}
\sigma_p^2(\bm w) = &~252 \sum_{{\ell_1}=1}^{L}\sum_{{\ell_2}=1}^{L} w_{\ell_1} w_{\ell_2} \text{cov}(\hat{R}_{\ell_1},\hat{R}_{\ell_2}),
\end{align}
where $252$ denotes 252 trading days per year, $\hat{R}_{\ell_1}$ and $\hat{R}_{\ell_2}$ denote log returns of the ${\ell_1}$-th and ${\ell_2}$-th stocks, respectively.

The second portfolio allocation method is based on the minimum variance portfolio, which provides investors with the lowest risk. The weights are determined by
\begin{align}
\label{con_mv1}
\bm{\hat{w}} =&~ \arg\min_{\bm w} {\sigma_{p}(\bm w)},\\
\text{s.t.}&~~\sum_{{\ell}=1}^{L} w_{\ell} = 1,~w_{\ell} \in [0,1],~{\ell}=1,...,L.\nonumber
\end{align}

\subsection{Portfolio Return Results}



\begin{figure}[!tb]
	\centering
	\subfloat{\includegraphics[width=3.0in]{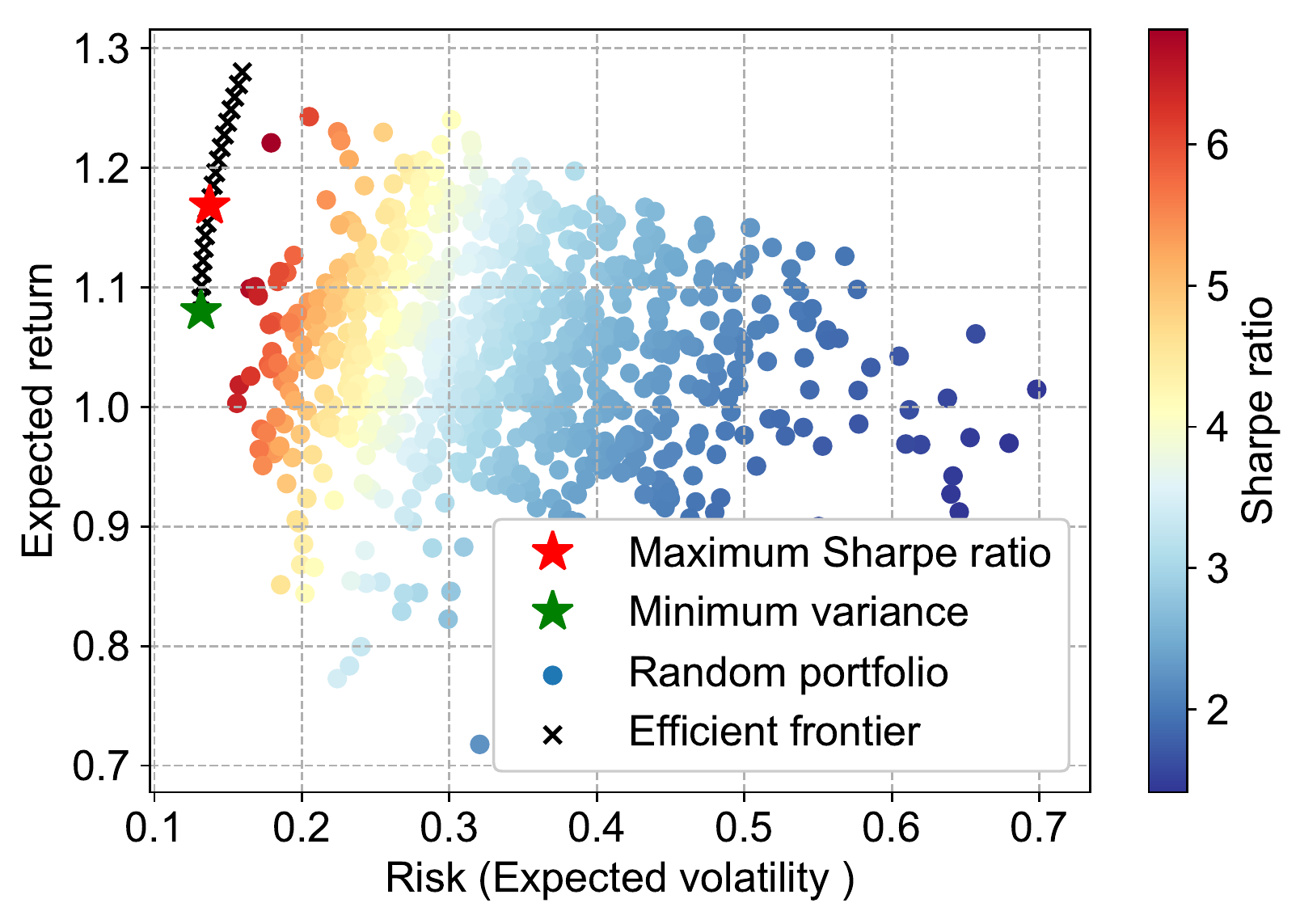}}
	\caption{Mid-LSTM portfolio allocation: 09/28/2018-12/24/2018}
	\label{figure:portfolio}
\end{figure}

The portfolio allocation analysis based on Mid-LSTM prediction is shown in Figure~\ref{figure:portfolio}. The red star on the curve is the mean-variance result and the green star is the minimum-variance result. We can see that these two portfolio allocation methods are suitable for midterm investment.

Table~\ref{tab:cumu} shows the portfolio return of the first set of asset. We can see that for both mean variance and minimum variance portfolio allocation methods, the average return based on the Mid-LSTM is better than that based on traditional methods. In Table~\ref{tab:cumu2}, we compare the portfolio return and Sharpe ratio of the second set of asset. We can see that when investing only in high market-related stocks, the average return and the Sharpe ratio based on the Mid-LSTM are both significantly higher than that based on traditional methods, which means that investment based on Mid-LSTM can achieve higher returns and lower risks.

\textcolor{black}{Note that the purpose of this empirical study is not to create a new portfolio allocation method, but rather to show the benefit of our new prediction approach. With a more accurate prediction of stock price, the already established properties among different portfolio allocation methods (mean-variance portfolio is the optimal one in terms of Sharpe Ratio in Figure~\ref{figure:portfolio}) should not be changed. Other PA methods with sub-optimal Sharpe Ratio will not beat current results.}

Furthermore, it is noteworthy that the proposed Mid-LSTM has good explanatory meanings to quantitative analyst. Analysts need to consider the impacts of various variables when making investment decisions. The weights of the linear regression model in Mid-LSTM can help analysts to understand the impacts of hidden variables in the market, making decisions more correct and less risky. \textcolor{black}{Moreover, the investment risk can be further reduced when selecting the stocks that are highly correlated with the market, since the experiment results show that the proposed Mid-LSTM has better anomaly circumvent performance when the stock is more correlated with the market.}

\section{Conclusions}

In this paper, \textcolor{black}{we bridge the deep neural network with the famous financial models (CAPM and ARMA), taking into consideration the market index and volume in stock price prediction to reduce the investment risk. We first propose a Mid-ARMA model} to represent the midterm stock price, which incorporates influential variables (volume and market) based on the ARMA model and CAPM model. Then, a Mid-LSTM deep neural network is proposed to predict midterm stock price according to the Mid-ARMA model, which combines the LSTM, hidden Markov model and linear regression networks. Experiment results based on the S\&P 500 stocks show that the proposed Mid-LSTM network can predict \textcolor{black}{the midterm stock price accurately even in the event of anomalies}, especially for those stocks that are high market-related. Portfolio return results show that the investment return can be significantly improved based on the predicted midterm prices of our Mid-LSTM network.

\bibliographystyle{ACM-Reference-Format}
\bibliography{sample_base}


\end{document}